\documentclass[twocolumn,superscriptaddress]{revtex4}
\pdfoutput=1

\usepackage{graphicx,color,url}
\definecolor{b}{rgb}{0,0,1.0}
\definecolor{r}{rgb}{1,0,0}

\begin{document}

\newcommand{\SZFKI}{Research Institute for Solid State Physics and Optics,
 P.O. Box 49, H-1525 Budapest, Hungary}

\newcommand{\LANL}{Condensed Matter and Thermal Physics $\&$ Center 
for Nonlinear Studies, Los Alamos National Lab, New Mexico 87545, USA}

\newcommand{\DAMTP}
{Department of Applied Mathematics and Theoretical Physics, University of Cambridge, UK}

\newcommand{\Tr}{\ensuremath{\textrm{Tr}}}

\title{Patterns in Flowing Sand: Understanding the Physics of Granular Flow}

\author{Tam\'as B\"orzs\"onyi}

\email{btamas@szfki.hu}
\affiliation{\SZFKI}
\affiliation{\LANL}
\author{Robert E. Ecke}
\affiliation{\LANL}

\author{Jim N. McElwaine}
\affiliation{\DAMTP}

% Pacs: 
% 47.57.Gc Granular flow
% 45.70.-n Granular systems
% 45.70.Qj Pattern formation

\begin{abstract}
Dense granular flows are often unstable and form inhomogeneous structures.
Although significant advances have been recently made in understanding 
simple flows, instabilities of such flows are often not understood. 
We present experimental and numerical results that show the formation
of longitudinal stripes that arise from instability of the uniform 
flowing state of granular media on a rough inclined plane. 
The form of the stripes depends critically on the mean density of the 
flow with a robust form of stripes at high density that consists of fast
sliding plug-like regions (stripes) on top of highly agitated 
\emph{boiling} material --- a configuration reminiscent of the 
Leidenfrost effect when a droplet of liquid lifted by 
its vapor is hovering above a hot surface.
\end{abstract}
\maketitle

\noindent Granular flows are ubiquitous in the environment and in
industry but there are still no known equations for general granular
systems. For flow on an inclined plane, however, progress has been
made~\cite{ba1954,sa1979,gdrmidi2004,ca2006}, and current
theories give a reasonable explanation of uniform flowing states. The
simplicity of the inclined-plane geometry plays a crucial role in such
descriptions because boundary effects, which can be exceedingly
complicated in granular flows, are well defined.  
This system is also well suited to Discrete Element Method
(DEM) simulations~\cite{sier2001,delo2007,tari2007} because it can be
treated periodically and steady flows often result, so good statistics
can be obtained. The combination of detailed
simulations and experiments has led to a solid understanding of steady
flow states in this system where experiments and simulations can be
well summarized~\cite{jofo2006,gdrmidi2004,arts2008,tari2007} by a
recently proposed model~\cite{jofo2006}. Little is known, however,
about the instabilities of this model or indeed of most granular
flows, and the model has only been well tested in simple shear states.
In contrast, the Navier-Stokes equation of fluid flow has been known
for over a century, and its accuracy has been repeatedly tested by
comparing the results of linear and weakly-nonlinear stability
analysis to experimental systems displaying an instability from a
simple state to one with a distinct pattern~\cite{crho1993}. Such an
approach, that is, investigating the growth of instabilities from
their respective steady states~\cite{fopo2001,prpa2000},
will certainly be very useful in testing granular constitutive models
and will provide critical tests for emerging theories of granular
flow.

The steady and fully developed state of a rapid, dilute granular flow
on a rough inclined plane was
shown experimentally to be unstable to the formation of longitudinal
vortices observed as lateral stripes~\cite{fopo2001}.  In this pattern
the downstream velocity and the layer height vary periodically across
the flow consisting of higher-slower and lower-faster regions.
The development is attributed to a mechanism analogous to the
Rayleigh-B\'enard instability in heated liquid layers~\cite{sphu1969}.
The average packing fraction  $\eta_{av}$ of this flow 
was below 0.3 corresponding to a relative density
$\rho_r = \eta_{av}/\eta_s \approx 0.5$, where $\eta_s$ is the static
packing fraction ($\eta_s$ is in between the random-closed-packed 
and random-loose-packed packing fractions \cite{bema1960}). This low 
density state, however, is hard to find either numerically or for some 
materials experimentally. 
Instead, we show that when increasing the plane inclination angle, the 
stripe state that emerges naturally is an instability of a {\it dense} 
uniform flow state, that this stripe state is robust and easy 
to find, and 
that the maxima of the downstream surface velocity correspond to the 
highest regions of the modulated height profile in qualitative agreement 
with the flow rule for the uniform state~\cite{po1999}.

\begin{figure}[htb]
  \includegraphics[width=\columnwidth]{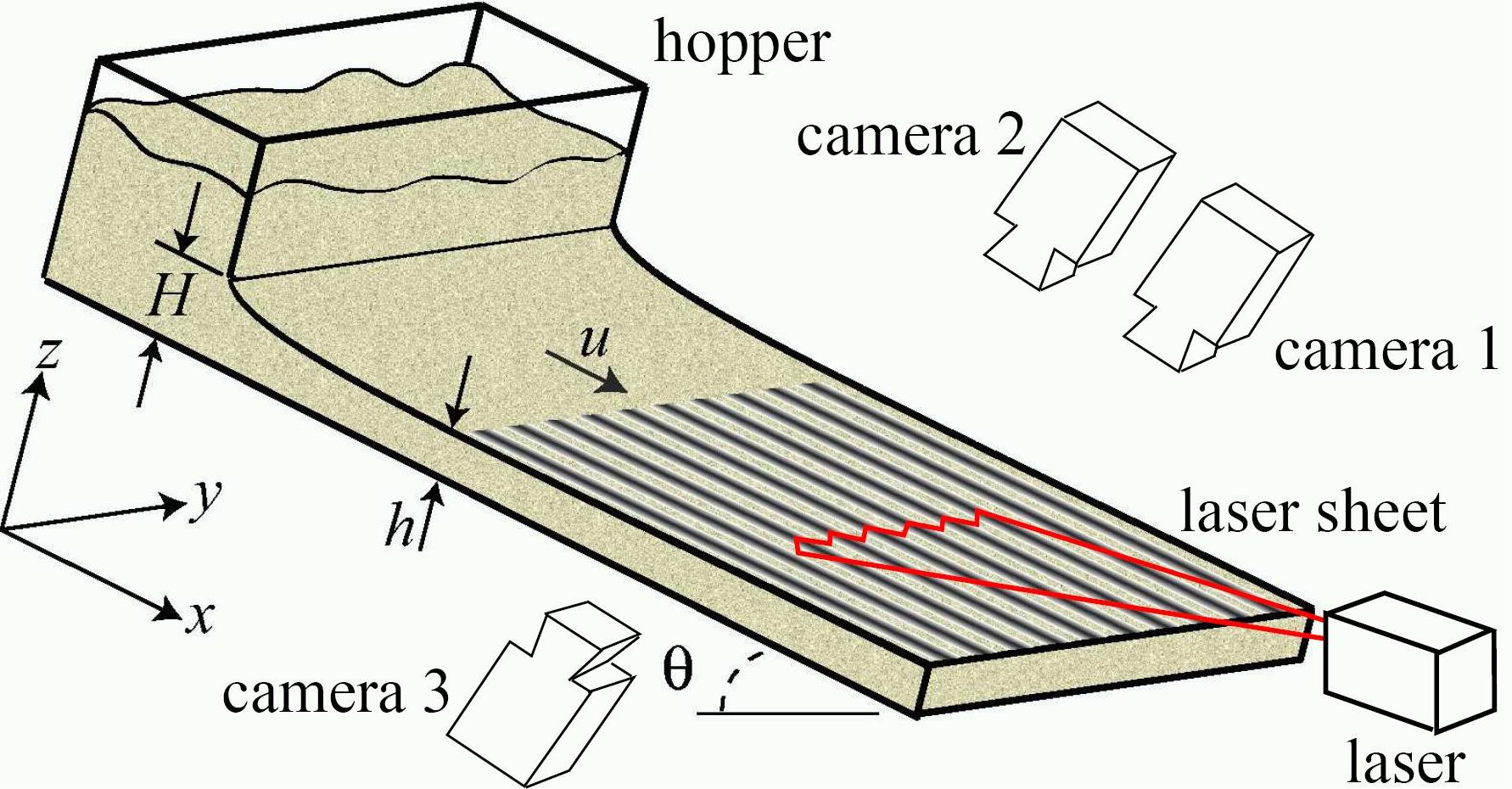}
  \caption{Schematic illustration of the experimental setup.}
  \label{setup}
\end{figure}

The flow was analyzed for a wide range of granular
materials including different sized sand and glass beads and
various copper samples with different particle shape.  We
demonstrate, using the apparatus illustrated in Fig.~\ref{setup}, that
longitudinal stripes are robustly observed over a broad range of flow
conditions with relative densities in the range $0.2 < \rho_r < 0.95$
(corresponding to about $0.12 < \eta_{av} < 0.57$), 
separated into the dense flow state for $\rho_r \gtrsim 0.6$ 
(i.e. $\eta_{av} \gtrsim 0.36$) and the 
dilute stripes for lower $\rho_r$ as described below.

In the present study two experimental setups were used. The
first apparatus, described in detail elsewhere~\cite{boec2006}, was
used to characterize flow regimes over a wide range of $\theta$.
Because the system was enclosed in a cylindrical tube, precise
measurements of the height profile $h(y)$ and determination of the
flow structure were performed in a second setup. It
consisted of a glass plate with dimensions
$2.27$\,m\,$\times\,0.4$\,m inclined at an angle $\theta=41.3^\circ$,
see Fig.~\ref{setup}. One layer of the $d=0.4$\,mm sand grains was
glued onto the surface of the glass plate in both setups to provide a
random roughness.  Measurements were repeated using sandpaper with
characteristic roughness of $0.19$\,mm (grit 80) and produced very
similar results. 
The surface velocity of the flow, with downstream and transverse
components $u^s$ and $v^s$, and the height profile $h(y)$, determined
by a laser sheet, were obtained simultaneously using two cameras
(Fig.~\ref{setup}). Using camera 3, the velocity at the bottom of
the layer ($u^b$ and $v^b$) was taken at a location where the glass
plate was clean. The flow velocities were determined using particle
image velocimetry on image sequences taken at 2000 frames/s. The 
relative density $\rho_r$ was measured using a method described in 
detail elsewhere~\cite{boec2006}. The majority of the data presented 
in this paper was obtained with sorted sand with $d=0.4\pm0.05$\,mm 
and $d=0.2\pm 0.05$\,mm. The stripe state was also detected for glass 
beads with $d=0.18\pm 0.05$\,mm, $d=0.36\pm 0.05$\,mm, and for four 
sets of copper particles with similar size ($d=0.16\pm0.03$\,mm) 
but various shapes. The angle of repose $\theta_r$ varied between 
$20.9^\circ \leq \theta_r \leq 33.8^\circ$ for the materials tested.  

DEM simulations were also performed to
investigate the instability. A soft particle model was used with a
damped linear spring for the normal force (coefficient of restitution
0.8) and Coulomb friction for the tangential force (coefficient of
friction 0.5). Particle stiffness was chosen so that the maximum
overlap was less than 1\%. The time step was 1/10 of the binary
collision time. The base was made of identical particles held at fixed
positions taken from another simulation where a thick layer was
allowed to form randomly. All quantities were non-dimensionalized
using the particle diameters and gravity. The instability was found
over a range of parameter values: slope angle 34--39$^\circ$,
restitution 0.80--0.95 and width greater than 50. Below we present results
from one typical simulation with a slope angle $37^\circ$. 
If the slope angle in the simulations is reduced below $\theta_r$ then
the flows come to rest with a packing fraction of 0.6. 
We use this packing fraction to normalize the results. The system
was periodic in the $x$ direction (down-slope length 24.3) and the $y$
direction (cross-slope width 120.15). The number of particles simulated 
was 55 761, so the volume of the particles over the $xy$ area corresponded 
to a height of 10 (height at 100\% packing fraction). The system was run
for several months until steady state was achieved. 
Contact stresses were calculated according to the method
of~\cite{gogo2002} using a Dirac delta function as the weight
function. The stresses, particle positions, particle velocities (first
and second moments), were gridded with spacing 0.1 vertically and
0.45 laterally using linear interpolation. These were calculated every 
program step and averaged over 500 time units.

\begin{figure}[ht]
  \includegraphics[width=\columnwidth]{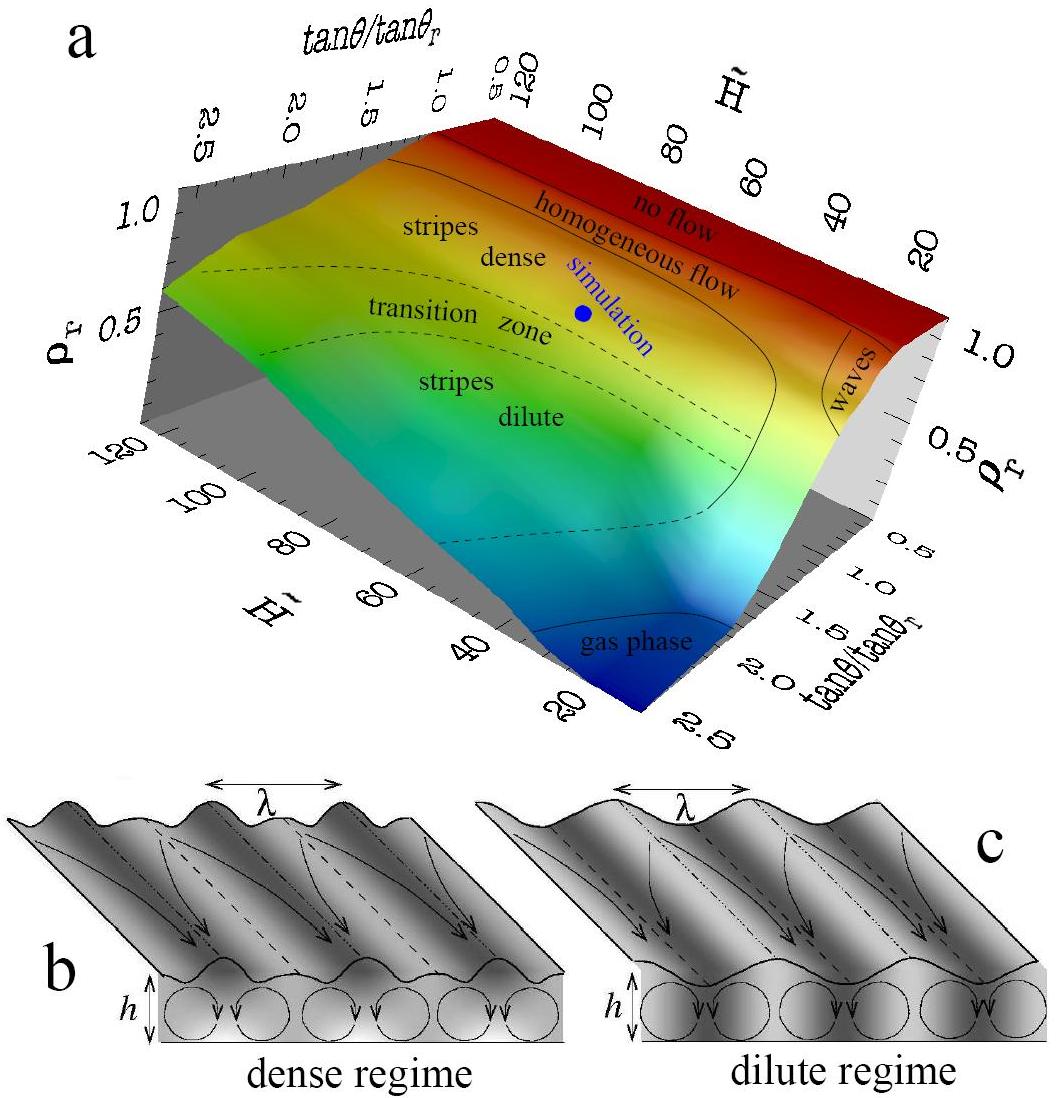}
  \caption{(a) The phase diagram of the system presenting the  
   mean flow density $\rho_r$ as a function of normalized plane 
   inclination $\tan \theta/\tan\theta_r$  and the normalized hopper 
   opening $\tilde{H}=H/d$ based on data taken for sand, glass beads 
   and various copper samples. 
   Various flow regimes are indicated, the bullet in the dense stripe
   domain corresponds to the simulation data presented.  
   Illustration of the flow structure in (b) dense and 
   (c) dilute regimes, gray levels indicate local density. }
  \label{density}
\end{figure}

\begin{figure}[ht]
  \includegraphics[width=\columnwidth]{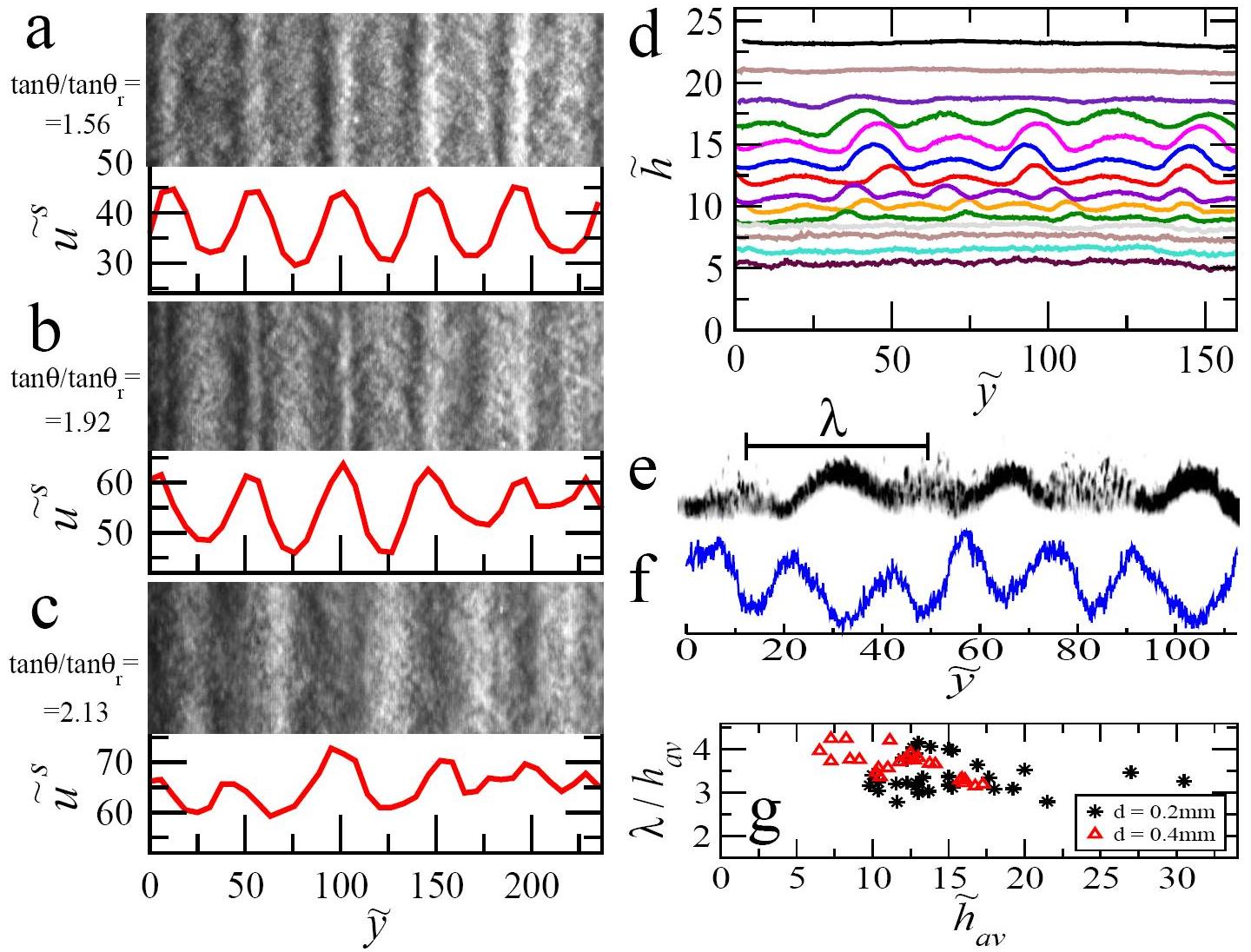}
  \caption{Images of the flow taken in reflected light 
    (illumination from the right) and normalized downstream 
    surface velocity $\tilde{u}^s=u^s/\sqrt{gd}$ as a 
    function of the normalized transverse coordinate 
    $\tilde{y}=y/d$ for sand with $d = 0.2$\,mm and 
    downstream distance from the outlet $x = 1.55$ m 
    at plane inclinations
    (a) $\theta = 42.6^\circ$;
    (b) $48.5^\circ$ and
    (c) $52.2^\circ$, corresponding to 
        $\tan \theta/\tan \theta_r = 1.56$,
        1.92 and 2.19, respectively.
   Data obtained at $x=2.13$\,m for sand with 
    $d=0.4$\,mm at $\theta = 41.3^\circ$: 
    (d) height profiles $h(y)$ taken at various hopper openings H, 
    (e) laser-line intensity (exposure time $4$\,ms), 
    (f) transmitted light intensity, and 
    (g) dimensionless wavelength $\lambda/h_{av}$ of 
    the pattern as a function of the normalized 
    mean flow thickness $\tilde{h}_{av}=h_{av}/d$. 
    To adjust $h_{av}$ the hopper opening $H$ was varied. 
  }
  \label{data}
\end{figure}

\begin{figure}[ht]
 \includegraphics[width=\columnwidth]{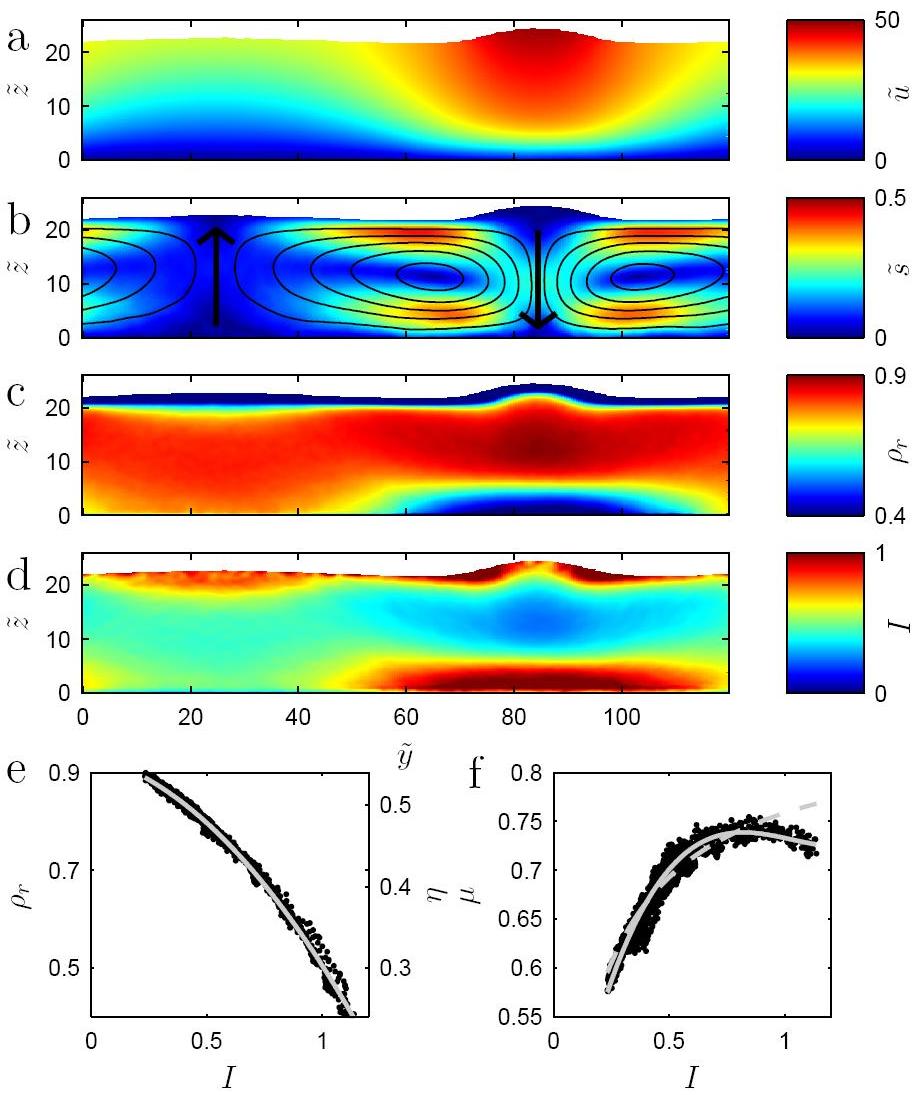}
 \caption{Results of the DEM simulations: (a) cross section of the 
   downstream velocity $\tilde{u}$, (b) speed in the
   $yz$ plane ($\tilde{s}=\sqrt{\tilde{v}^2+\tilde{w}^2}$) with streamlines, 
   (c) relative density $\rho_r=\eta/0.6$ or packing fraction $\eta$,
   (d) the inertial 
   number I, (e) dependence of relative density on $I$ and best
   quadratic fit, and (f) dependence of effective friction $\mu$ on
   the inertial number $I$. Solid line is best cubic fit. Dashed line
   is best fit to Pouliquen model (Eq.2 in \cite{jofo2006}).}
 \label{MD}
\end{figure}

In the experiments the stripe state for all the materials tested has 
qualitatively similar characteristics, and there is no sharp transition 
between the states observed in the dense and dilute flow regimes. 
Nevertheless, the flow structure of the dense flow state is quite 
different from the dilute flow case reported earlier~\cite{fopo2001}. 
The structure of the dense stripe state consists of
relatively-narrow, dense, fast-moving regions that are also the
highest. In the dilute regime the fast moving region corresponds to a
height minimum, as schematically illustrated in Fig.~\ref{density}b,c.
In both cases, however, the grains sink in the fastest moving regions.
The continuous transition between the two regimes is characterized by
increasing height of the slow moving region as the plane inclination
is increased as illustrated in Fig.~\ref{data}a-c or in movies
taken for various materials at~\cite{movie}. The density
decreases with increasing $\theta$ in a similar way for all materials
as shown in Fig.~\ref{density}a, where $\rho_r$ is plotted as a
function of the normalized hopper opening $\tilde{H}=H/d$ and normalized 
plane inclination $\tan \theta/\tan \theta_r$. Generally, stripes are only
observed for $\tan\theta/\tan\theta_r > 1.25$. Stripes with the
structure typical for the dense regime are observed for $0.6<\rho_r<0.95$ 
(corresponding to $0.36<\eta_{av}<0.57$), whereas the 
dilute regime, when it is observed, exists in the range for 
$0.2<\rho_r<0.7$ (i.e.\ $0.12<\eta_{av}<0.42$).  
In the following, we characterize the stripe structure in the
dense regime using experimental data shown in Fig.~\ref{data} obtained
for sand with $d=0.4$\,mm and $d=0.2$\,mm and numerical results shown
in Fig.~\ref{MD} obtained from DEM simulations.

The flow pattern~\cite{movie} has a downstream surface velocity
$u^s(y)$ and a lateral surface velocity $v^s(y)$. The downstream
velocity has a relatively large modulation of
$(u^s_{\mathrm{max}}-u^s_{av})/u^s_{av} \leq 0.2$, whereas the lateral
velocity is very slow with maximal value $v^s_{\mathrm{max}} \leq
0.04\, u^s_{av}$ where $u^s_{av}$ denotes the average downstream
surface velocity. Cross sections of the velocity of the fully
developed state in simulations are shown in Fig.~\ref{MD}a,b showing
very similar downstream velocity modulation, but somewhat smaller
lateral velocities. In the experiments the lateral velocity measured
at the surface $v^s(y)$ and at the bottom of the layer $v^b(y)$, is of
opposite direction in accord with the flow structure obtained from the
simulations (see Fig.~\ref{MD}b).

The experimentally observed height variation in Fig.~\ref{data}d
agrees nicely with the simulation results (Fig.~\ref{MD}a-d). The fast
stripes correspond to a higher narrow maximum of the $h(y)$ curve but
another set of less pronounced maxima is present between them. Thus,
instead of a sinusoidal $h(y)$ profile observed for the dilute regime,
a more complex $h(y)$ is seen where a higher, global maximum
corresponds to the fast flowing region and a lower, local maximum
corresponds to the slower region. The double peak is also seen in the
transmitted light intensity in Fig.~\ref{data}f, but the wavelength
of other important measures of the pattern, e.g., the downstream velocity 
or velocity in the $yz$ plane 
(see Fig.~\ref{data}a-c and Fig.~\ref{MD}a-b) do not change 
during the dilute-dense transition so the emergence of a 
nonsinusoidal height profile does not correspond to a full 
frequency-doubling. The pattern is observed only in a 
finite range of
the flow thickness with the strongest amplitude at $10 < \tilde{h} <
18$.  The wavelength of the pattern $\lambda$ is related to the 
mean flow thickness $h_{av}$ as $2.8 < \lambda/h_{av} < 4.5$ as shown 
in Fig.~\ref{data}g, so the cross section of a roll is 
elongated as compared to the traditional Rayleigh-B\'enard rolls with 
nearly circular cross section \cite{crho1993}.

Numerical simulations enable us to visualize spatial variations of the
relative density, Fig.~\ref{MD}c, and of the inertial number $I$ in
Fig.~\ref{MD}d. The inertial number, usually defined for incompressible 
flows~\cite{jofo2006}, can be extended for
the compressible case. We define $I = {d \sqrt{\rho}|D'|}/{\sqrt{p}}$,
where $D'$ is the deviatoric strain tensor
($D'=D-\Tr(D)/3$), $\rho$ the density, $p$ the pressure, and we use
the norm $|A|=\sqrt{\Tr{(AA^T)/2}}$. We do not consider normal
pressure differences and define $p=-\Tr(\sigma)/3$,
where $\sigma$ is the stress tensor and $\sigma'=\sigma+p$. This is
equivalent to the Pouliquen definition when the flow is incompressible
($\Tr(D)=0$).
The inertial number is proportional to the shear rate and to the ratio
of the collisional stress to the total stress. The flow has the highest
density in the fast moving region where $I$ 
(and shear rate) is lowest. We identify this region as a ``plug'' sliding
fast on top of a ``boiling'' region with very low relative density and
high inertial number and shear rate, a configuration reminiscent of the 
Leidenfrost effect~\cite{le1756} when a droplet of liquid lifted by its 
vapor is hovering above a hot surface.  
Experimental data visualizing the
level of fluidization at the surface (Fig.~\ref{data}e) and the
profile of the transmitted light intensity (Fig.~\ref{data}f) fully
agree with this picture.

The relative density (or packing fraction) determined from the 
simulation
data decreases
monotonically with increasing shear rate, see Fig.~\ref{MD}e, and shows
an amazing collapse over a wide range of densities and values of the
inertial number $I$. This suggests that, at least for fast chute flows, a
simple equation of state giving the pressure as a function of shear
rate, and density is possible.  To test the rheology we calculate the
effective friction coefficient $\mu$ by $\mu = - \Tr(\sigma' D')/{p|D'|}$.
This is the $\mu$ that minimizes the residual error $\left|\sigma'+\mu
  \frac{D'}{|D'|}\right|$. Simpler calculations of $\mu$, e.g., 
$\mu=-\sigma_{xz}/\sigma_{zz}$ or $\mu=-\sigma_{xz}/p$, produce poor
results (no collapse would be seen in Fig.~\ref{MD}f), due to the
complicated strain field. 
This definition is an extension of the 
Pouliquen model~\cite{jofo2006} to include compressible flows.
Fig.~\ref{MD}f shows the effective friction coefficient as a function of
$I$ and demonstrates a reasonably good collapse. The
data does not fit well with Pouliquen rheology and is much more
strongly curved and appears to even decrease for large $I$ 
(Fig.~\ref{MD}f).  At low shear rates $\mu$ increases with increasing
$I$, but at $I=0.7$ there appears to be a turnover above which $\mu$
decreases with increasing $I$. We believe that this behavior of the
system is a key feature leading to the instability. Namely, by
increasing the flow thickness above a certain value, the inertial
number near the plane reaches a threshold above which the effective
friction starts decreasing. As a consequence the local inertial number
(shear rate) grows even more, leading to stronger fluidization near the
plane.  At the same time the fluidization drops in the upper layer (a
plug develops) and this plug slides even faster on top of the expanded
fluidized region. This mechanism agrees with other results for chute
flows~\cite{tari2007}, but in that case the simulations were too
narrow for the lateral instability to develop. On the surface the plug
absorbs material from the two sides as the surface fluidization is
larger in that region, so the plug grows. In the simulations, as the
instability develops the mean flow speed decreases until a new lower
velocity is reached at which point the instability has saturated.
Thus, the instability may play an important role on steeper slopes,
where no simple, steady state is expected, by increasing the effective
viscosity. Though we have framed our discussion in terms of 
$\mu$, it is difficult to draw too many conclusions from these results
for a Pouliquen rheology. Our simulation data shows density 
differences, normal stress difference and that the deviatoric stress 
tensor is not aligned with the strain tensor. These all disagree with
the assumptions of the Pouliquen rheology indicating that a 
considerably more complicated rheology is necessary along with
an equation of state to describe these flows.

This system provides a very interesting case for studying granular
rheologies because there is a complicated strain and stress field but
very simple boundary conditions. Since the flow is steady, accurate
measurements of all the flow variables in a simulation are possible,
the only constraint being computer time. This system is therefore
ideal for developing and validating granular theories in new ways. An
intriguing possibility is that the lateral ridges and furrows observed 
in large rock avalanches~\cite{kelfoun2008} may be the result of the 
same instability.

This work was funded by the US Department of Energy under
Contract No.\ DE-AC52-06NA25396. Authors benefited from discussions
with I.S.\ Aranson, W.B.\ Daniel, O.\ Pouliquen, M.K.\ Rivera, E.\
Somfai and M.\ van Hecke. T.B. was supported by the Bolyai
J\'anos research program, and the Hungarian Scientific Research Fund
Grant No. OTKA-F60157. J.N.M. was supported by the Engineering
and Physical Sciences Research Council (UK).

\end{document}